\documentclass[doublecol]{epl2}
\usepackage{amsfonts,amssymb,amsmath,hyperref,graphicx,epsfig}
\newcommand{\ket}[1]{| #1 \rangle}

\newcommand{\avg}[1]{\langle #1 \rangle}

\title{The Sagnac Effect in Optical Lattices with Laser-Assisted Tunneling}
\author{Bo-Nan Jiang\inst{1,2}\thanks{E-mail: \email{bnjiang@siom.ac.cn}} \and Xiao-Gang Wei\inst{1,2}\thanks{E-mail: \email{xggwei@hotmail.com}} \and Guo-Wan Zhang\inst{1,2} \and Jia-Hua Li\inst{1,2} \and Yong-Jie Cheng\inst{1,2} \and Cheng Xu\inst{1,2}}
\shortauthor{B.-N. Jiang \etal}
\institute{
  \inst{1} Quantum Engineering Research Center, China Aerospace Science and Technology Corporation, Beijing 100854, China\\
  \inst{2} Beijing Institute of Aerospace Control Devices, Beijing 100854, China
}

\pacs{07.60.Ly}{Interferometers}
\pacs{03.75.-b}{Matter waves}
\pacs{37.10.Jk}{Atoms in optical lattices}

\abstract{
We propose a scheme to realize rotation sensing through the use of optical lattices with laser-assisted tunneling. We theoretically demonstrate that competition between the rotation and the spin-orbit coupling governs the spin-dependent response of the cyclotron dynamics of the spin-orbit coupled bosons. The Sagnac-type cumulative phase can be read out from the envelope of a beat-frequency time evolution of the population imbalance in the spin-balanced system and enhanced by cyclotron motion. We also theoretically show that the sensitivity limit of the spin-orbit-coupled system to rotational motion can reach $4\times10^{-7} \mathrm{rad} s^{-1} Hz^{1/2}$.}

\begin{document}
\maketitle
\section{Introduction}
Over the past decades, the Sagnac effect has given birth to an important class of inertial sensors employing the rotation-dependent phase shift $\Phi=\frac{4\pi\Omega A}{\lambda v}$ between two counterpropagating laser light or matter waves \cite{Chow1985,Stedman1995,Vali1976,Pritchard1997,Kasevich1997}. In the Sagnac geometry \cite{Sagnac1913}, $\Omega$ is the rotation, $A$ is the area enclosed by the Sagnac loop, $\lambda$ is the wavelength of the laser light or matter waves, and $v$ is the corresponding velocity \cite{Storey1994}. Owing to the rapid developments in laser cooling and manipulation of neutral atoms \cite{Nobel1997,Nobel2001,Bloch2012}, the application of atom interferometers to precision rotation measurements has been explored \cite{Cronin2009,Pritchard1997,Kasevich1997}, with the projected sensitivity of an atom-based Mach-Zehnder gyroscope demonstrated to exceed that of an optical gyroscope by a factor of $\sim10^{11}$ \cite{Dowling1993,Dowling1998}. Although suffering a lower atom flux and a smaller enclosed loop area, the atom interferometer gyroscope can potentially outperforms the photon-based system by several orders of magnitude \cite{Kasevich1997}.

However, because only a half circuit is established around the full-loop Sagnac geometry, the Mach-Zehnder-type atom interferometer achieves only half of the phase shift originally considered by Sagnac \cite{Sagnac1913,Kasevich1997,Gustavson2000}. The phase shift or sensitivity of the Mach-Zehnder gyroscope can of course be increased considerably by having the laser light or matter wave travel more than a half Sagnac loop. In a photon-based system, this has been achieved by guiding two counterpropagating laser light through an optical fiber that circles the Sagnac loop many times \cite{Vali1976}. Thus, naturally, we ask whether an atom-based system featuring counterorbiting matter-wave components can also sense rotation.

One such atom-based system has recently been realized in optical lattices with laser-assisted tunneling (LAT) \cite{Bloch2013,Ketterle2013,Bloch2015,Zoller2015}, where opposite spin components of the spin-orbit-coupled bosons manifest cyclotron motion with opposite chiralities on the edge \cite{Bloch2013,Kennedy2013}. However, to answer the question above, further theoretical efforts are needed for two reasons. First, the competition between rotation and spin-orbit coupling has not yet been discussed, either experimentally nor theoretically \cite{Spielman2013}. Second, the readout of the cumulative phase therein is also a challenging task, since there is no interference in analogy to the ring fiber interferometer between two bosonic currents with distinct spins.

To address these issues, we theoretically study a rotating spin-orbit-coupled bosonic system in this work. We demonstrate that the opposite spin components of the bosons in optical lattices with LAT are respectively accelerated and decelerated by the rotation, leading to a Sagnac-type cumulative phase enhanced by the cyclotron motion. Furthermore, we show that competition between the rotation and the spin-orbit coupling dramatically generates a beat-frequency time evolution of the population imbalance in the spin-balanced system, from the envelope of which the cumulative phase (or the rotation) can be read out. We also theoretically estimate that the sensitivity limit of the spin-orbit-coupled system to the rotational motion is $4\times10^{-7} \mathrm{rad} s^{-1} Hz^{1/2}$.

\begin{figure}[tbp]
\onefigure[width=0.8\linewidth]{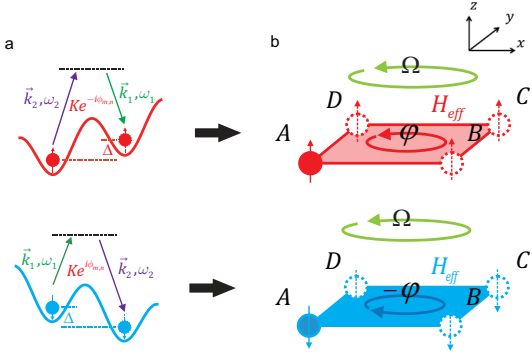}
\caption{(Color online) (a) Laser-assisted tunneling for (top) $\ket{\uparrow}$ and (bottom) $\ket{\downarrow}$ bosons. (b) Schematic diagram of a rotating spin-orbit-coupled bosonic system.
\label{fig:model}}
\end{figure}

\section{Model}
As shown in Fig. \ref{fig:model}, we consider noninteracting bosonic $^{87} $Rb$ $ atoms trapped in a two-dimensional optical lattice, where the opposite spin states $\ket{\uparrow}$ and $\ket{\downarrow}$ denote the Zeeman states $\ket{F=1,m_F=-1}$ and $\ket{F=2,m_F=-1}$ respectively, and $\sigma_z$ is the corresponding Pauli spin matrix \cite{Bloch2013}. Along $x$, the normal tunneling of the atom is suppressed by a magnetic field gradient $\Delta\sigma_z$ in this direction. The spin-dependent resonant LAT is reestablished with two far-detuned Raman laser beams of frequency detuning $\delta\omega=(\omega_2-\omega_1)\sigma_z=\frac{\Delta}{\hbar}\sigma_z$ and momentum transfer $\delta\vec{k}=(\vec{k}_2-\vec{k}_1)\sigma_z$, realizing the spin-orbit coupling \cite{Kolovsky2011,Creffield2013}. The spin-orbit-coupled bosonic $^{87} $Rb$ $ atoms corotate with the setup with an angular velocity $\Omega$ along $z$ \cite{Spielman2009}. By generalizing the treatments in Refs. \cite{Ketterle2013,Jaksch2003,Tino2012,Miyake2013,Bhat2007,Moller2010,Sachdeva2010}, we derive, in rotating-frame coordinates, a modified Harper Hamiltonian that models the rotating spin-orbit-coupled bosonic system under the tight-binding approximation:
\begin{eqnarray}
\label{eq:Heff}
H_{eff}=&-&\sum_{m,n}[Ke^{-i\phi^{so}_{m,n}-i\phi^{ro,x}_{m,n}}\sum_\sigma\hat{b}^\dag_{m+1,n,\sigma}\hat{b}_{m,n,\sigma} \nonumber \\
    &+&Je^{-i\phi^{ro,y}_{m,n}}\sum_\sigma\hat{b}^\dag_{m,n+1,\sigma}\hat{b}_{m,n,\sigma}+h.c.]  \nonumber \\
    &+&\sum_{m,n,\sigma}V_T(m,n)\hat{b}^\dag_{m,n,\sigma}\hat{b}_{m,n,\sigma},
\end{eqnarray}
where $\hat{b}_{m,n,\sigma}$ annihilates a boson in the spin state $\ket{\sigma}$ at site $(m,n)$. $K$ and $J$ are the effective tunneling amplitudes of the atom along $x$ and $y$, and we set $K/\hbar=0.27\times2\pi\mathrm{kHz}$ and $J/\hbar=0.53\times2\pi\mathrm{kHz}$ following the experimental situation \cite{Bloch2013}. $V_T(m,n)=\frac{1}{2}M(\omega_h^2-\Omega^2)\vec{r}^2_{m,n}$ is a trapping potential generated by a weak external harmonic confinement with a trapping frequency $\omega_h \approx 25\times2\pi\mathrm{Hz}$ and the centrifugal potential \cite{Bloch2014}, where $M$ is the mass of the $^{87} $Rb$ $ atom.

Here, we discuss the effect of trapping potential $V_T(m,n)$ on the cyclotron dynamics of the rotating spin-orbit-coupled bosons. First, on an elementary plaquette, since $V_T(m,n)\ll J,K$, the effect of $V_T(m,n)$ can be reasonably neglected. Second, to reach a large Sagnac loop in optical lattices, a toroidal optical dipole trap is applied to the current setup through the use of red-detuned Laguerre-Gaussian beams along $z$ \revision{\cite{Zoller2015,Wright2000}}. The spin-orbit-coupled bosons trapped in the toroidal confinement travel around a ring lattice and experience a constant $V_T(m,n)$ at each site because of the symmetry of traps. Thus, $V_T(m,n)$ can be treated as a trivial constant in Eq. (\ref{eq:Heff}), and the demonstration on the elementary plaquette can be straightforwardly generalized.

The competition between rotation and spin-orbit coupling conceives in the spatially varying phase of the effective tunneling in Eq. (\ref{eq:Heff}). The spin-orbit coupling that takes the form $V_{so}\sigma_z(xp_y-yp_x)$ arises from the spatially varying phase $\phi^{so}_{m,n}=\varphi(m+n)\sigma_z$, leading to the quantum spin Hall effect (QSHE) \cite{Bloch2013,Kennedy2013,Bernevig2006}. In the time-reversal-symmetric QSHE system, the $\ket{\uparrow}$ and $\ket{\downarrow}$ bosons manifest cyclotron orbits with opposite chiralities and the cumulative phase on the edge $\Phi(\Box)=\sum\limits_{\Box,\sigma}\phi^{so}_{m,n}=0$ \cite{Bloch2013}. The rotation, on the other hand, destroys the QSHE by breaking the time-reversal symmetry. The corresponding non-time-reversal-symmetric interaction that takes the form $V_{ro}(xp_y-yp_x)$ arises from the spatially varying phases $\phi^{ro,x}_{m,n}=-\frac{M}{\hbar}\int^{\vec{r}_{m+1,n}}_{\vec{r}_{m,n}}\vec{\Omega}\times\vec{r}\cdot d\vec{r}=\frac{M\Omega a^2n}{\hbar}$ and $\phi^{ro,y}_{m,n}=-\frac{M}{\hbar}\int^{\vec{r}_{m,n+1}}_{\vec{r}_{m,n}}\vec{\Omega}\times\vec{r}\cdot d\vec{r}=-\frac{M\Omega a^2m}{\hbar}$. The competition between rotation and spin-orbit coupling leads to a Sagnac-type cumulative phase $\Phi(\Box)=\sum\limits_{\Box,\sigma}(\phi^{so}_{m,n}+\phi^{ro,x}_{m,n}+\phi^{ro,y}_{m,n})=N_{loop}\frac{4M\Omega A}{\hbar}$ \cite{Kasevich1997,Gustavson2000}, thus governing the underlying physics of the rotation sensing. $V_{so}$ and $V_{ro}$ denote the strength of interactions, $a$ is the lattice constant. Since the Sagnac-type cumulative phase increases with the number of the cyclotron loops $N_{loop}$, the spin-orbit-coupled bosonic system can be treated as an ultracold-atom analog of the fiber ring interferometer, but with a different readout procedure, which shall be demonstrated in the following sections.

\begin{figure*}[tbp]
\onefigure[width=0.7\textwidth]{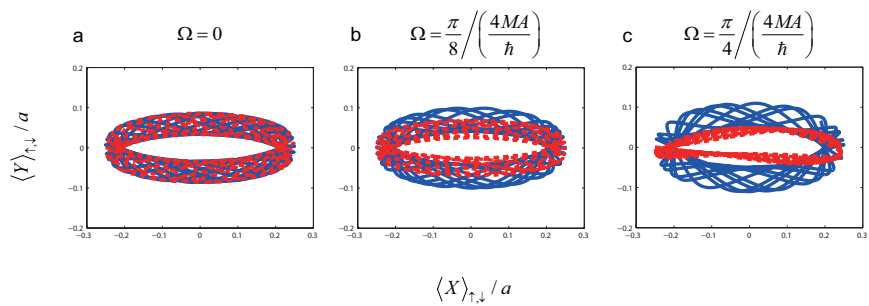}
\caption{(Color online) The cyclotron orbits of the $\ket{\uparrow}$ (blue solid) and $\ket{\downarrow}$ (red dashed) mobile bosons in the ($a$) nonrotating and ($b-c$) rotating spin-orbit-coupled systems in a period of $T=20\mathrm{ms}$, with increasing strengths of rotation, $\Omega=0$, $\frac{\pi}{8}/\frac{4MA}{\hbar}$, and $\frac{\pi}{4}/\frac{4MA}{\hbar}$. The total atom number is $N=16$, and $\varphi=\pi/4$.}
\label{fig:orbits}
\end{figure*}

\section{The Sagnac effect on a plaquette}
\revision{Considering the Sagnac-type cumulative phase discussed above, nontrivial sensitivity to the rotational motion can be reached with spin-orbit-coupled bosons in a sufficiently large optical ring lattice. The underlying physics of such a multi-site system can be transparently revealed by a minimal model - a four-site modified Harper Hamiltonian on an elementary plaquette \cite{Bloch2013}. Thus, here, the competition between rotation and spin-orbit coupling are monitored by the cyclotron dynamics of $N$ $\ket{\uparrow}$ and $\ket{\downarrow}$ bosons on a plaquette isolated in a two-dimensional optical lattice.} We study the cyclotron dynamics assuming an spin-balanced initial state
\begin{equation}
    \ket{\Psi_0}=\frac{1}{\sqrt{N!}}[\frac{1}{2}(b^{\dag}_{A,\uparrow}+b^{\dag}_{D,\uparrow}+b^{\dag}_{A,\downarrow}+b^{\dag}_{D,\downarrow})]^N\ket{0}
\end{equation}
describing $N$ bosons prepared in a superposition of two spin components on sites $A$ and $D$, where $\ket{0}$ denotes the vacuum state and sites $\{A,B,C,D\}=\{(0,0),(1,0),(1,1),(0,1)\}$ \cite{Li2013}. The physics described is robust against the choice of the initial state. The quantities characterizing the cyclotron motion are the time-dependent mean atom positions of the $\ket{\uparrow}$ and $\ket{\downarrow}$ bosons along $x$ and $y$, which are defined as
\begin{eqnarray}
    \avg{X}_\sigma &=&(N_{B,\sigma}+N_{C,\sigma}-N_{A,\sigma}-N_{D,\sigma})a/2N \nonumber \\
    \avg{Y}_\sigma &=&(N_{C,\sigma}+N_{D,\sigma}-N_{A,\sigma}-N_{B,\sigma})a/2N,
\end{eqnarray}
where $N_{m,n,\sigma}=\avg{\hat{b}^\dag_{m,n,\sigma}\hat{b}_{m,n,\sigma}}$. By solving the Schr\"{o}dinger equation with Eq. (\ref{eq:Heff}), we obtain the trajectories of the spin-orbit-coupled bosons. In the nonrotating QSHE system (Fig. \ref{fig:orbits}$a$), the counterorbiting $\ket{\uparrow}$ and $\ket{\downarrow}$ bosons undampedly travel around the loop of a same area $A$. With the rotation increasing from 0 to $\Omega=\frac{\pi}{4}/\frac{4MA}{\hbar}$ (Figs. \ref{fig:orbits}$b-c$), the cyclotron orbit of the $\ket{\uparrow}$ bosons expands with the increasing cumulative phase $\Phi_\uparrow(\Box)=N_{loop}[\varphi+\frac{2M\Omega A}{\hbar}]$, while the $\ket{\downarrow}$ bosons manifests the cyclotron motion that contracts with the decreasing cumulative phase $\Phi_\downarrow(\Box)=-N_{loop}[\varphi-\frac{2M\Omega A}{\hbar}]$. The expansion and contraction of the cyclotron orbits indicate that the rotation accelerates and decelerates the motion of the $\ket{\uparrow}$ and $\ket{\downarrow}$ bosons respectively, being consistent with the broken time-reversal symmetry.

\begin{figure}[tbp]
\onefigure[width=0.4\textwidth]{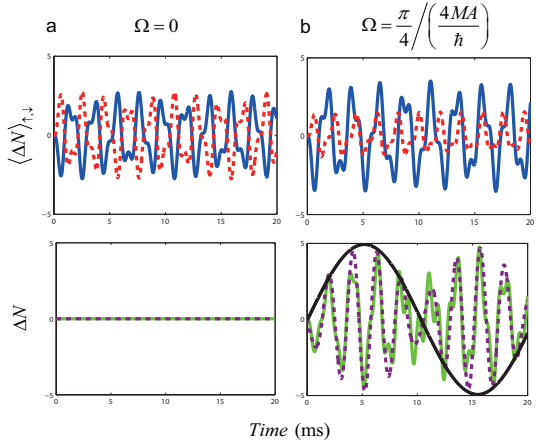}
\caption{(Color online) The particle current in the ($a$) nonrotating and ($b$) rotating spin-orbit-coupled systems with balanced spin populations. The population imbalance $\Delta N$ (green solid line) modulated by an envelope of frequency $\omega_{beat}/2$ (black solid line) is well described by a beat-frequency function $\Delta N_{beat}$ (purple dashed line). $\Delta N_{\uparrow}$ and $\Delta N_{\downarrow}$ are also plotted with blue solid and red dashed lines respectively. The total atom number is $N=16$, and $\varphi=\pi/4$.}
\label{fig:beats}
\end{figure}

\begin{figure}[tbp]
\onefigure[width=0.4\textwidth]{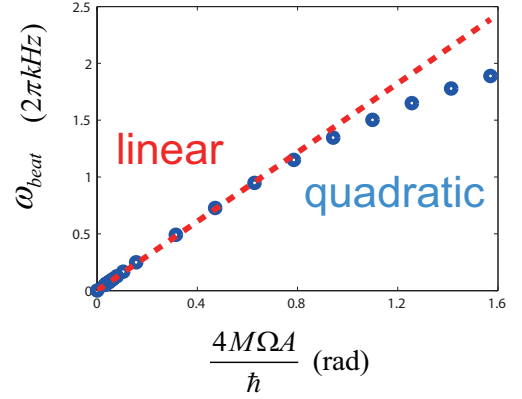}
\caption{(Color online) The beat frequency of $\Delta N$ as a function of the parameter $\frac{4M\Omega A}{\hbar}$. The total atom number is $N=16$, and $\varphi=\pi/4$.}
\label{fig:std}
\end{figure}

To obtain a quantitative relation between the cumulative phase (or the rotation) and the cyclotron dynamics, we further study the particle current through the time evolution of the population imbalance defined as
\begin{equation}
    \Delta N=\Delta N_{\uparrow}+\Delta N_{\downarrow}=N_{C}+N_{D}-N_{A}-N_{B},
\end{equation}
where $N_{m,n}=N_{m,n,\uparrow}+N_{m,n,\downarrow}$. In the nonrotating spin-balanced QSHE system, the time evolution of $\Delta N_{\uparrow}$ and $\Delta N_{\downarrow}$ are perfectly mirrored \cite{Bloch2013}, leading to a time-independent population imbalance $\Delta N=0$ (Fig. \ref{fig:beats}$a$). Thus, the particle current can be fully neglected. As the rotation respectively accelerates and decelerates the cyclotron motion of the $\ket{\uparrow}$ and $\ket{\downarrow}$ bosons (Fig. \ref{fig:orbits}), the time evolution of $\Delta N_{\uparrow}$ and $\Delta N_{\downarrow}$ manifest a mismatch of both the oscillation amplitude and the frequency (Fig. \ref{fig:beats}$b$, top), resulting in a time-dependent population imbalance and dramatically leading to a measurable particle current in the spin-balanced system (Fig. \ref{fig:beats}$b$, bottom). For an analytically understanding of the nontrivial particle current in the rotating spin-orbit-coupled system, we approximately treat the time evolution of $\Delta N_{\uparrow}$ and $\Delta N_{\downarrow}$ as sinusoidal functions
\begin{eqnarray}
    \Delta N_\uparrow &\approx& A_\uparrow\sin{\omega_\uparrow t} \nonumber \\
    \Delta N_\downarrow &\approx& A_\downarrow\sin{\omega_\downarrow t},
\end{eqnarray}
where $A_{\uparrow,\downarrow}$ and $\omega_{\uparrow,\downarrow}$ are the oscillation amplitude and the frequency. Straightforwardly, we obtain a beat-frequency fit function that approximately describes the time evolution of population imbalance as
\begin{eqnarray}
\label{eq:fit}
    \Delta N_{beat}&=&\Delta N_\uparrow+\Delta N_\downarrow \nonumber \\
     &\approx& (A_\uparrow+A_\downarrow)\sin{\frac{\omega_{beat}}{2} t}\cos(\frac{\omega}{2} t) \nonumber \\
    &+&(A_\uparrow-A_\downarrow)\cos{\frac{\omega_{beat}}{2} t}\sin(\frac{\omega}{2} t),
\end{eqnarray}
where the beat frequency $\omega_{beat}=\omega_\uparrow-\omega_\downarrow$ and $\omega=\omega_\uparrow+\omega_\downarrow$. It is shown in Fig. \ref{fig:beats} that $\Delta N_{beat}$ well captures the underlying physics of the particle current found in the numerics: ($a$) In the spin-balanced QSHE system, the particle current with $\omega_{beat}=0$ is not measurable, which is consistent with the experimental observation of the oscillation amplitude being proportional to the spin polarization \cite{Bloch2013}. ($b$) As the rotation destroys the QSHE, the time evolution of the population imbalance exhibits fast oscillations with frequency $\omega/2$, modulated by an envelope $(A_\uparrow+A_\downarrow)\sin{\frac{\omega_{beat}}{2} t}+(A_\uparrow-A_\downarrow)\cos{\frac{\omega_{beat}}{2} t}$ of frequency $\omega_{beat}/2$, indicating that the competition between the rotation and the spin-orbit coupling generates an effective temporal spin polarization oscillating with the beat frequency in the spin-balanced system. Comparing the above two points, one can realize that the beat-frequency time evolution of the population imbalance is an unique phenomenon in the rotating spin-orbit-coupled system. By applying the beat-frequency fit function in Eq. (\ref{eq:fit}), we derive in Fig. \ref{fig:std} a quantitative relation between the cumulative phase (or the rotation) and $\omega_{beat}$ from the exact numerical simulations. As the rotation increases, the beat frequency increases linearly with the parameter $\frac{4M\Omega A}{\hbar}$ in the weak rotation regime and goes quadratic in the strong rotation regime, demonstrating that the beat frequency is an important measurement of the competition between the rotation and the spin-orbit coupling, and the Sagnac-type cumulative phase (or the rotation) can be read out from the envelope of the time evolution of the population imbalance. One can also easily derive that this envelope increases to the maximum at $t_{max}=\frac{2}{\omega_{beat}}\arctan{\frac{A_\uparrow+A_\downarrow}{A_\uparrow-A_\downarrow}}$, when the corresponding cumulative phase $\Phi(\Box)=N_{loop}(t_{max})\frac{4M\Omega A}{\hbar}\sim\frac{\pi}{2}$. It means that, in analogy to the fiber ring interferometer, the readout signal of a weak rotation can be increased considerably after $N_{loop}(t_{max})$ cyclotron loops.

\section{Experimental considerations}
From the abovementioned demonstrations, the quantitative relation between the the rotation and the beat frequency obviously plays a key role in the rotation sensing using spin-orbit-coupled bosons. The rotation sensing is roughly achieved by experimentally measuring the time evolution of the population imbalance and numerically fitting the corresponding beat frequency with Eq. (\ref{eq:fit}). Any variation of the rotation will affect the beat frequency following the the quantitative relation shown in Fig. \ref{fig:std}. From the experimental side, the time evolution of the population imbalance can be probed by non-destructive phase-contrast imaging that is extensively used for in-situ density measurement of bosonic gases \cite{Ketterle1997,Greiner2009,Chin2009}. The image of the density distribution at arbitrary time $t$ is obtained by collecting photons scattered coherently from the cloud of bosons \cite{Greiner2009}, and the corresponding population imbalance is straightforwardly derived from the density distribution. We theoretically estimate the sensitivity of the spin-orbit-coupled system to rotational motion using the number-phase uncertainty principle applied in Refs. \cite{Dowling1993,Dowling1998,Dowling2012}. And we derive a sensitivity limit from the uncertainty of the phase-contrast imaging as
\begin{eqnarray}
    \Omega_{min}=\frac{\hbar}{N_{loop}(t)\sqrt{N_{sc}}(4MA)},
\end{eqnarray}
where $N_{sc}$ is the flux of scattered photons. Assuming we realize a ring lattice with diameter $d \approx 22a$ by adding red-detuned Laguerre-Gaussian beams to a square optical lattice, and load $N=10^6$ spin-orbit-coupled $^{87} $Rb$ $ atoms. Using a typical collected number of photons per atom per second $2\times10^4$ for the in-situ imaging system \cite{Greiner2009}, the detector receives a flux of scattered photons $N_{sc}\sim2\times10^{10}s^{-1}$. Considering the bosons accomplish one loop in $\sim2 ms$ on the plaquette, $N_{loop}(t)\sim 29$ per second in the ring lattice. Thus, the numerical value for the minimal rotation measurable is calculated to be $\Omega_{min}\sim4\times10^{-7} \mathrm{rad} s^{-1} Hz^{1/2}$ for $t=1s$, which is very close to the best result $10^{-9} \mathrm{rad} s^{-1} Hz^{1/2}$ reported by the state-of-the-art atom-based Mach-Zehnder gyroscopes \cite{Arnold2006,Gupta2005}.

One unique avenue to improve the sensitivity limit of the spin-orbit-coupled system is through the interorbital spin-exchange interaction \cite{Ye2014,Bloch2014a,Inguscio2014}. In an earlier work \cite{Jiang2014}, we demonstrated that the strong spin-exchange interaction between $^1$S$_0$ and $^3$P$_0$ clock states of the alkaline-earth-metal atoms in optical lattices with LAT restores the QSHE of the Kondo singlet. As the interorbital spin-exchange interaction mimicking the Kondo-exchange interaction \cite{Gorshkov,Rey2010}, the effective mass of the quasiparticle could be increased by several orders of magnitude as in the heavy-fermion materials, thus leading to a considerable improvement in the sensitivity limit.

\section{Conclusion}
We studied the Sagnac effect in a spin-orbit-coupled bosonic system through cyclotron dynamics on a plaquette. We showed that the competition between rotation and spin-orbit coupling generates a measurable beat-frequency time evolution of the population imbalance, which is completely distinguishable from a Quantum spin Hall effect system with no spin polarization. We also demonstrated that a Sagnac-type cumulative phase can be read out from the envelope of the particle current and considerably increased by cyclotron motion. We further estimated that the sensitivity limit of the spin-orbit-coupled system can reach $4\times10^{-7} \mathrm{rad} s^{-1} Hz^{1/2}$. We hope our work could provide a reasonable guideline for the experimental investigation of the Sagnac effect in optical lattices with laser-assisted tunneling in the future.

\acknowledgments
We thank Colin Kennedy for helpful discussions. This work was supported by the National Natural Science Foundation of China (Grant No. 11304007, Grant No. 11204011, and Grant No. 60907031).

\end{document}